\documentclass[runningheads]{llncs}
\usepackage[T1]{fontenc}
\usepackage{graphicx}%
\usepackage{multirow}%
\usepackage{amsmath,amssymb,amsfonts}%
\usepackage{mathrsfs}%
\usepackage[title]{appendix}%
\usepackage{xcolor}%
\usepackage{textcomp}%
\usepackage{manyfoot}%
\usepackage{booktabs}%
\usepackage{algorithm}%
\usepackage{algorithmicx}%
\usepackage{algpseudocode}%
\usepackage{listings}%
\usepackage{braket}
\usepackage{caption}
\usepackage{subcaption}
\usepackage{printlen}
\usepackage{url}
\usepackage{orcidlink}
\usepackage{bbding}
\begin{document}
\title{The Role of Quantum in Hybrid Quantum-Classical Neural Networks: An Empirical Assessment}
\titlerunning{The Role of Quantum in Hybrid Quantum-Classical Neural Networks}
%
\author{
Dominik~Freinberger\,\Envelope\,\orcidID{0009-0006-0545-2083} \and
Philipp~Moser\orcidID{0000-0002-9717-6197}
}

\authorrunning{D. Freinberger and P. Moser}

\institute{
Research Unit Medical Informatics, RISC Software GmbH,\\
Softwarepark 32a, 4232~Hagenberg, Upper Austria, Austria\\
\email{dominik.freinberger@risc-software.at}
}
\maketitle              
\begin{abstract}
Quantum machine learning has emerged as a promising application domain for near-term quantum hardware, particularly through hybrid quantum-classical models that leverage both classical and quantum processing. Although numerous hybrid architectures have been proposed and demonstrated successfully on benchmark tasks, a significant open question remains regarding the specific contribution of quantum components to the overall performance of these models. In this work, we aim to shed light on the impact of quantum processing within hybrid quantum-classical neural network architectures through a rigorous statistical study under ideal, noise-free simulation. We systematically assess a selection of commonly used HQNN architectures on three binary medical classification tasks covering medical signal data as well as planar and volumetric images. In doing so, we examine the influence attributable to classical and quantum aspects such as encoding schemes, entanglement, and circuit size. Within this experimental setting, we find that in best-case scenarios, hybrid models show performance comparable to their classical counterparts. However, in most cases, performance metrics deteriorate when the investigated QNN components replace matched classical processing layers. Our multi-modal analysis provides empirical insights into the contributions of the tested quantum components and advocates for cautious claims and design choices for hybrid models in near-term applications.

\end{abstract}
%
%
%
\section{Introduction}
\label{sec:introduction}

Quantum machine learning (QML) has generated considerable excitement as a potential pathway to leverage near-term quantum devices for practical applications \cite{cerezo_challenges_2022,wang_comprehensive_2024}. Quantum neural networks (QNNs), parameterized quantum circuits executed on quantum computers and trained by classical hardware \cite{mitarai_quantum_2018}, have emerged as particularly promising models due to a certain resilience to noise \cite{sharma_noise_2020} and their structural similarities with classical neural networks (NNs)~\cite{perez-salinas_data_2020}. Despite these appealing features, current implementations of QNNs face critical limitations, primarily due to hardware constraints. Particularly severe is the limited number of qubits available on existing quantum devices (or simulatable classically) and the error accumulation as well as the barren plateau phenomenon in deep circuits\cite{mcclean_barren_2018}, which restrict the ability to encode high-dimensional data directly into QNNs.

To address these constraints, hybrid quantum-classical neural networks (HQNNs) have been proposed. HQNNs integrate QNNs into conventional NN architectures and are frequently explored for their promise of combining the strengths of both paradigms. NNs often act as initial feature extractors, producing fewer but highly informative latent features to be encoded into the QNN. The QNN is claimed to offer benefits including improved expressivity and training ability \cite{abbas_power_2021}, among others. Despite the widespread adoption of HQNNs and their successful demonstrations on benchmark \cite{mari_transfer_2020,matic_quantum-classical_2022} and real-world tasks \cite{rainjonneau_quantum_2023,sagingalieva_hybrid_2023,xiang_quantum_2024}, a central question remains largely unaddressed or even overlooked: To what extent does the quantum component truly contribute to the overall hybrid model performance? 

Recent studies have offered mixed results. Some literature suggested that specific hybrid quantum-classical models offer performance improvements, such as enhanced accuracy~\cite{hafeez_h-qnn_2024,perelshtein_practical_2022,xiang_quantum_2024,zeng_multi-classification_2022} particularly with fewer training data \cite{sagingalieva_hybrid_2023}. Other works remained more cautious, presenting hybrid models as a promising approach requiring further exploration and validation \cite{mari_transfer_2020,matic_quantum-classical_2022}. Yet others have taken a more critical look at the contribution of quantum components in specific settings. For example, \cite{kolle_disentangling_2024} compared a hybrid architecture that encodes features from a pre-trained classical encoder into a QNN with angle encoding, evaluating it against various classical baselines and other hybrid models with pre-trained classical components. They reported that the inclusion of quantum components did not yield improvements over fully classical models. Similarly, as part of a broader QML benchmark, \cite{bowles_better_2024} investigated the transformations induced by a particular HQNN architecture and found them to be qualitatively similar to those produced by a corresponding classical model.

Motivated by this gap and the lack of consensus in the current understanding, our study expands upon these studies by providing a systematic and rigorous statistical analysis to assess the impact of quantum processing within a family of commonly used HQNN architectures. In contrast to prior work such as \cite{kolle_disentangling_2024}, we focus on fully hybrid training schemes where classical and quantum components are optimized jointly, rather than relying on pre-trained classical encoders or separate training stages. By systematically varying classical pre-processing complexity, latent space dimensionality, quantum encoding methods, and measurement strategies, we aim to compare the contributions of quantum versus purely classical components on the overall model score. We evaluate hybrid models against fully classical counterparts across three distinct medical data modalities: One-dimensional ECG signals, two-dimensional breast ultrasound images and three-dimensional chest CT image data, which represent some of the most frequent data types, especially in healthcare applications.

The remainder of the paper is organized as follows. Section~\ref{sec:methods} details our methodological setup, including the datasets, the HQNN architecture as well as variations of the hybrid and classical models and the training and evaluation protocols. Section~\ref{sec:results} presents our empirical findings across all configurations and datasets. Finally, Section~\ref{sec:discussion} discusses the implications of our results and concludes the paper.
%
%
%

\section{Methods}
\label{sec:methods}

\subsection{Datasets and Modalities}
\label{sec:data-modalities}
\begin{figure}[ht]
  \centering
  \begin{subfigure}[b]{0.28\columnwidth}
      \centering
      \includegraphics[width=\textwidth]{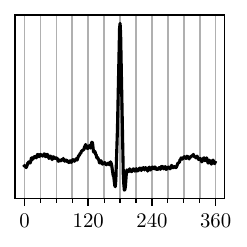}
      \captionsetup{aboveskip=-4pt}
      \caption{MIT-BIH}
  \end{subfigure}
  \hfill
  \begin{subfigure}[b]{0.24\columnwidth}
      \centering
      \includegraphics[width=\textwidth]{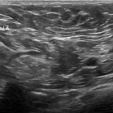}
      \caption{BreastMNIST}
  \end{subfigure}
  \hfill
  \begin{subfigure}[b]{0.24\columnwidth}
      \centering
      \includegraphics[width=\textwidth]{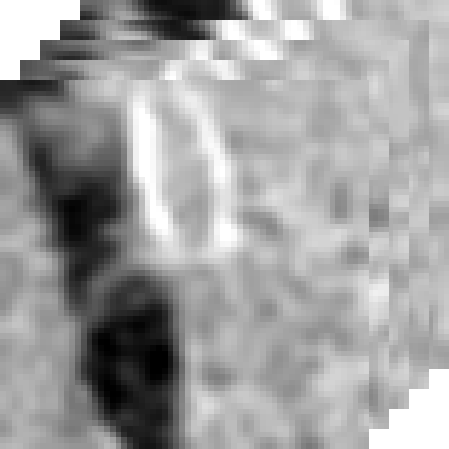}
      \caption{NoduleMNIST}
  \end{subfigure}
  \caption{\textbf{Examples from the three datasets used in this study:} 1D ECG (a), 2D breast ultrasound (b), and 3D chest CT (c).}
  \label{fig:data-samples}
\end{figure}

We focused on binary classification tasks across diverse real-world medical data modalities using three publicly available datasets: the 1D MIT-BIH Arrhythmia dataset (electrocardiogram recordings) \cite{goldberger_physiobank_2000,moody_impact_2001}, the 2D BreastMNIST dataset (breast ultrasound images), and the 3D NoduleMNIST3D dataset (chest CT volumes), the latter two being part of the bigger MedMNIST collection \cite{yang_medmnist_2023}. Fig.~\ref{fig:data-samples} shows illustrative samples. The 1D MIT-BIH Arrhythmia dataset contains 105,026 annotated ECG recordings sampled at 360 Hz from 47 subjects. Each sample represents a 1-second cardiac cycle (360 features). We formulate a binary classification task distinguishing normal beats from arrhythmic episodes. The 2D BreastMNIST dataset comprises 780 grayscale breast ultrasound images labeled as benign or malignant, resized to $224 \times 224$ pixels with values normalized to [-1, 1]. For the 3D modality, we use NoduleMNIST3D, consisting of 1,633 CT scans ($64\times 64\times 64$ voxels) of pulmonary nodules classified as malignant or benign. No additional pre-processing was applied to this dataset.

We sampled 7,064 instances from the MIT-BIH dataset and created five cross-validation folds, ensuring no subject overlap between training and validation folds to prevent data leakage. Similarly, for the MedMNIST datasets we created five separate training and validation folds based on the provided training and validation split. All training and validation folds were balanced in terms of class distribution.

\begin{figure}[ht]
\centerline{\includegraphics[width=1.0\textwidth]{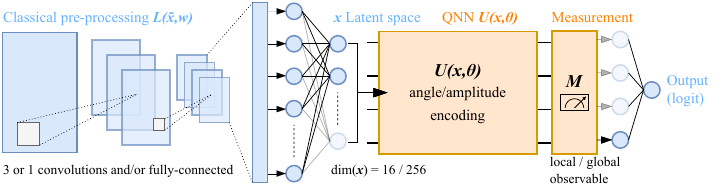}}
\caption{\textbf{Schematic illustration of the general HQNN architecture.} A classical network $L(\boldsymbol{\Tilde{x}}, \boldsymbol{w})$ (with 3, 1 or no convolutional layers) and a fully-connected layer (with or without $\pi \cdot \tanh$ activation) first extracts a latent feature vector $\boldsymbol{x}$ (16 or 256 features). $\boldsymbol{x}$ is then encoded into a quantum circuit $U(\boldsymbol{x}, \boldsymbol{\theta})$ using either angle or amplitude encoding (with optional entanglement). The expectation value of either a local or global Pauli‑Z measurement is finally linearly mapped to the output logit.}
\label{fig:hqnn}
\end{figure}

\subsection{Hybrid Model Architecture and Classical Reference}
\label{sec:model-architecture}

We adopted a prototypical hybrid model design inspired by~\cite{mari_transfer_2020}. The architecture comprised classical NN layers, $L(\boldsymbol{\Tilde{x}}, \boldsymbol{w})$, pre-processing the raw input $\boldsymbol{\Tilde{x}}$ into a latent feature representation $\boldsymbol{x}$. These features $\boldsymbol{x}$ were given to the QNN $U(\boldsymbol{x}, \boldsymbol{\theta})$ for further processing. The QNN output, obtained by measuring an observable $M$, was post-processed by a linear fully-connected layer yielding the final logits. Fig.~\ref{fig:hqnn} illustrates this design. This architecture supports various input/output dimensions without constraints by the QNNs encoding capacity. Further, we isolated the contribution of the quantum circuit, by removing classical pre-processing stages or replacing the QNN entirely with classical layers. We implemented three variants of $L$, to systematically assess the influence of classical pre-processing:

\begin{itemize}
    \item \textbf{3conv:} Three convolutional layers, each followed by Batch Normalization, ReLU activation, and Max-Pooling.
    \item \textbf{1conv:} A single convolutional layer with Batch Normalization, ReLU activation, and Max-Pooling.
    \item \textbf{0conv:} A single fully-connected layer without activation directly mapping the flattened input $\boldsymbol{\Tilde{x}}$ to the latent representation $\boldsymbol{x}$.
\end{itemize}

Convolutional and pooling operations were adapted to the input dimensionality. The output from \textbf{3conv} and \textbf{1conv} was flattened before a final fully-connected layer produced the latent feature vector $\boldsymbol{x}$. We explored two latent dimensions: \textbf{16} and \textbf{256}. For angle-based feature encoding, we compared using a scaled activation function $\pi \cdot \tanh$ (mapping features to $[-\pi, \pi]$), versus no activation before the quantum encoding. For the QNN $U$, we used 4 or 8 qubits for latent dimensions \textbf{16} or \textbf{256}, respectively. This pairing was chosen to match the logarithmic scaling of amplitude encoding ($\log_2(16)=4$ and $\log_2(256)=8$), allowing a controlled comparison between angle and amplitude encoding architectures at fixed circuit width. We investigated four distinct QNN architectures, two using angle encoding and two using amplitude encoding:
\begin{itemize}
    \item \textbf{Ang-RY:} The latent vector $\boldsymbol{x}$ was partitioned into segments matching the number of qubits. Each segment was embedded using angle encoding via RY-rotations. These embedding layers alternated with variational layers comprising arbitrary single-qubit rotations followed by CNOT gates in a strongly entangling circular pattern. A final variational layer was applied after the last embedding layer. Although all latent features were encoded only once in the circuit, classical layers could have implicitly learned re-uploading important features \cite{schuld_effect_2021}.
    
    \item \textbf{Ang-Arb:} Similar to \textbf{Ang-RY}, but used arbitrary single-qubit rotations $R(\phi, \theta, \omega)$ around all axes to encode three features per gate. Variational layers also used arbitrary single-qubit rotations, with entanglement via CZ gates in an alternating nearest-neighbor pattern (even-indexed pairs 0-1, 2-3... were entangled first, then odd-indexed pairs 1-2, 3-4... were entangled in the next layer). This design was inspired by \cite{perez-salinas_data_2020}, adapted such that the latent feature vector was encoded once in the circuit, relying on classical layers to potentially learn data re-uploading as in the $\textbf{Ang-RY}$ scheme. A final variational layer was applied.

    \item \textbf{Amp-Gen:} The normalized latent vector $\boldsymbol{x}$ was embedded using amplitude encoding in a single step. This was followed by a variational layer structure identical to that in \textbf{Ang-RY}, including the strongly entangling circular pattern of CNOT gates and the final variational layer. This ensured that the number of variational layers (and therefore trainable parameters) was equal to the \textbf{Ang-RY} architecture.
   
    \item \textbf{QCNN:} Latent features $\boldsymbol{x}$ were embedded using amplitude encoding. The \textbf{QCNN} architecture then utilized alternating quantum convolutional and pooling layers inspired by classical convolutional neural networks. Quantum convolution layers applied two-qubit unitary gates arranged in alternating pairings, first coupling even-indexed qubit pairs and then odd-indexed pairs with wrap-around connections. Quantum pooling layers reduced dimensionality by applying parameterized two-qubit unitary gates to pairs of qubits, effectively encoding the information into a single qubit per pair and subsequently disregarding one qubit, thus reducing the system's dimensionality. This procedure was repeated until a single output qubit remained, whose measurement determined the output \cite{cong_quantum_2019,qiskit_convnet}.    
\end{itemize}
Motivated by prior benchmarking results showing that removing entanglement can leave QML model performance unchanged or even improve it in some settings \cite{bowles_better_2024}, we compared circuits with and without entangling gates in the variational layers (except for \textbf{QCNN}, where entanglement is integral). The circuit output was the expectation value of one of two observables:
\begin{itemize}
    \item \textbf{Local:} Separate Pauli-Z expectation values for each qubit.
    \item \textbf{Global:} Tensor product of Pauli-Z on all qubits, yielding the global observable $\bigotimes_{i=1}^N Z_i$.
\end{itemize}
The \textbf{QCNN} inherently used a single Pauli-Z measurement on the final qubit. Variational parameters were initialized from $\mathcal{N}(0, (0.01\pi)^2)$.

Classical reference models used identical pre-processing $L(\boldsymbol{\Tilde{x}},\boldsymbol{w})$ (\textbf{3conv}, \textbf{1conv}, \textbf{0conv}) and latent dimensions (\textbf{16}, \textbf{256}) as described above for the hybrid models as well as a final linear output layer. In place of the QNN, we evaluated four classical processing variations:
\begin{itemize}
    \item \textbf{none:} Latent features were directly mapped to the output (via the linear post-processing layer).
    \item \textbf{fcnone:} A single hidden fully-connected layer (no activation).
    \item \textbf{fcrelu:} A single hidden fully-connected layer with ReLU activation.
    \item \textbf{mlp:} A multi-layer perceptron with three hidden fully-connected layers using ReLU activations.
\end{itemize}

The classical reference models were intentionally designed as matched architectural controls rather than as state-of-the-art task-specific baselines. They share pre-processing depth and latent dimensionality with the hybrid configurations, allowing us to isolate the marginal contribution of the QNN block under controlled conditions rather than to evaluate absolute performance ceilings on the underlying datasets.

\subsection{Training and Statistical Evaluation}
\label{sec:evaluation-setup} 
Models were trained using PyTorch \cite{paszke_pytorch_2019}. Quantum circuits were simulated using PennyLane \cite{bergholm_pennylane_2022} with the ideal, noise-free \texttt{default.qubit} simulator. We used the Adam optimizer with binary cross-entropy loss with logits, a batch size of \textbf{256} for the 1D dataset and 64 for the 2D and 3D datasets, and a learning rate of 0.001. We employed a 5-fold cross-validation strategy as described in Section~\ref{sec:data-modalities} for each model configuration and dataset. Performance was evaluated each epoch using ROC-AUC, average precision, and balanced accuracy. The best score for each metric within each fold was recorded, and results were averaged across the five folds to obtain a final score for each model configuration. Statistical significance was assessed using nonparametric tests (Wilcoxon signed-rank or Mann-Whitney U test, as appropriate) with Bonferroni correction for multiple comparisons, using a significance level of $\alpha = 0.05$.
%
%
%
\section{Results}
\label{sec:results}

\subsection{Influence of Classical Pre-Processing Layers}
\label{sec:influence-of-classical-pre-processing}
\begin{figure}[ht]
\centerline{\includegraphics[width=1.0\textwidth]{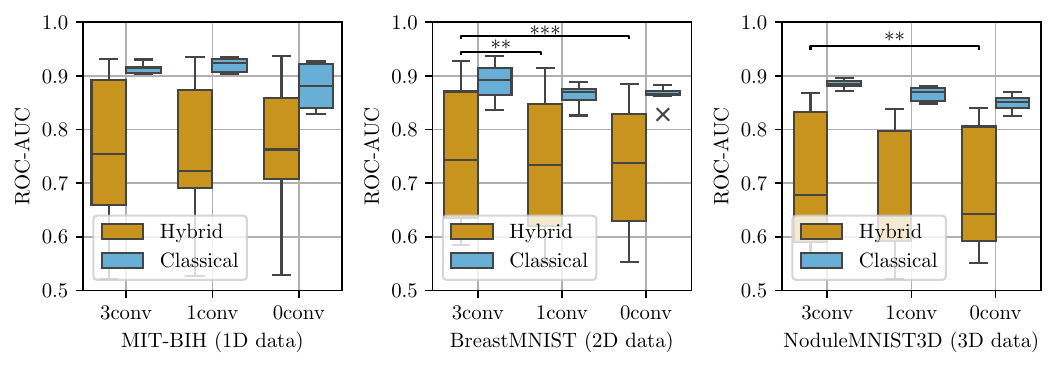}}
\caption{\textbf{Hybrid and classical model performances with different levels of classical pre-processing across data modalities.} The influence of the pre-processing depth (\textbf{3conv}, \textbf{1conv}, \textbf{0conv}) is statistically compared between hybrid models and classical reference models (*: $p<0.05$, **: $p<0.01$, ***: $p<0.001$).}
\label{fig:influence-of-classical-pre-processing}
\end{figure}

Across all data modalities, hybrid models exhibited considerably wider interquartile ranges in performance compared to purely classical models, often with significantly lower median ROC-AUC scores (Fig.~\ref{fig:influence-of-classical-pre-processing}). In the 1D and 2D datasets, the best-performing hybrid configurations achieved scores comparable to, or occasionally exceeding their classical counterparts (e.g., \textbf{Amp-Gen} with \textbf{0conv} pre-processing, latent dim. \textbf{16}, entanglement and \textbf{global} observable on the MIT-BIH data). Regarding the influence of classical pre-processing, we observed statistically significant differences between the \textbf{3conv} and \textbf{1conv} configurations $(p < 0.01)$ as well as between the \textbf{3conv} and \textbf{0conv} configurations $(p < 0.001)$ in the 2D setting. Similarly, in the 3D setting there was a significant difference $(p < 0.01)$ between the \textbf{3conv} and \textbf{0conv} setting. This trend was less pronounced in the 1D dataset, where no significant difference in hybrid model performance was observed between the classical pre-processing configurations.

\begin{figure}[ht!]
\centerline{\includegraphics[width=1.0\textwidth]{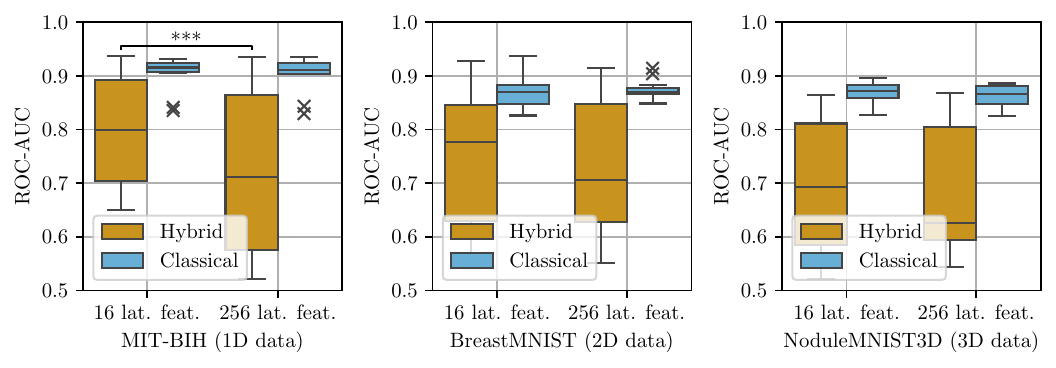}}
\caption{\textbf{Hybrid and classical model performances with different latent space dimensions across data modalities.} The influence of the latent space dimension (\textbf{16} vs. \textbf{256} features) is statistically compared between hybrid models and classical reference models (*: $p<0.05$, **: $p<0.01$, ***: $p<0.001$).}
\label{fig:latent_dim_boxplot}
\end{figure}

\subsection{Influence of Latent Space Dimension}
\label{sec:influence-of-latent-space-dimension}
We examined the impact of the latent feature space dimension on model performance in Fig.~\ref{fig:latent_dim_boxplot}. For hybrid models, the smaller latent dimension of \textbf{16} generally led to higher median ROC-AUC scores compared to a larger dimension of \textbf{256}. This difference was most pronounced and statistically significant in the 1D case ($p < 0.001$). Across all modalities and both latent dimensions, the purely classical models consistently achieved significantly higher median performance than the hybrid models ($p < 0.001$ in most comparisons). Among the classical models, the best results were obtained with \textbf{mlp} processing, \textbf{3conv} pre-processing, and \textbf{16} latent dim. for the 2D and 3D datasets, and with \textbf{fcrelu} processing, \textbf{1conv} pre-processing, and \textbf{256} latent dim. for the 1D dataset.

\begin{figure}[h]
\centerline{\includegraphics[width=.6\columnwidth]{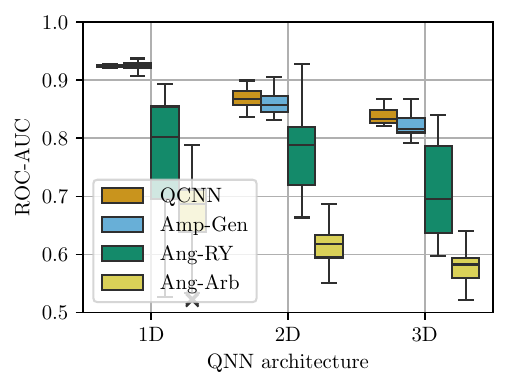}}
\caption{\textbf{ROC-AUC scores achieved by hybrid models, grouped by QNN architecture.} Amplitude encoding models (\textbf{Amp-Gen}, \textbf{QCNN}) generally outperformed angle encoding models (\textbf{Ang-RY}, \textbf{Ang-Arb}).}
\label{fig:qnns-boxplot}
\end{figure}

\subsection{QNN Architecture and Feature Encoding}
Among the QNN architectures, amplitude encoding (\textbf{Amp-Gen}, \textbf{QCNN}) resulted in significantly higher median scores across all datasets compared to angle encoding (\textbf{Ang-RY}, \textbf{Ang-Arb}), as shown in Fig.~\ref{fig:qnns-boxplot}. The \textbf{QCNN} architecture yielded the best median performance on 2D and 3D data. Notably, one specific configuration with angle encoding (\textbf{Ang-RY} with \textbf{3conv} pre-processing, \textbf{16} latent dim., entanglement and \textbf{global} observable) achieved the highest performance in the 2D dataset. The \textbf{Ang-Arb} architecture consistently performed the worst.

\subsection{Latent Layer Activation Function}
For angle encoding models (\textbf{Ang-RY}, \textbf{Ang-Arb}), omitting the scaling activation function ($\pi \cdot \tanh$) before quantum encoding led to significantly better performance across all datasets (Fig.~\ref{fig:hybrid_hparams_boxplot}, left panel).

\subsection{Entanglement in the Variational Circuit}
Comparing models with (\textbf{ent.}) versus without entangling (\textbf{no ent.}) operations in the variational circuit (excluding \textbf{QCNN}), we observed a slight, nonsignificant trend towards better median performance when entanglement was present across all data modalities (Fig.~\ref{fig:hybrid_hparams_boxplot}, middle panel).

\subsection{Type of Observable}
\begin{figure}[t!]
\centerline{\includegraphics[width=1.0\textwidth]{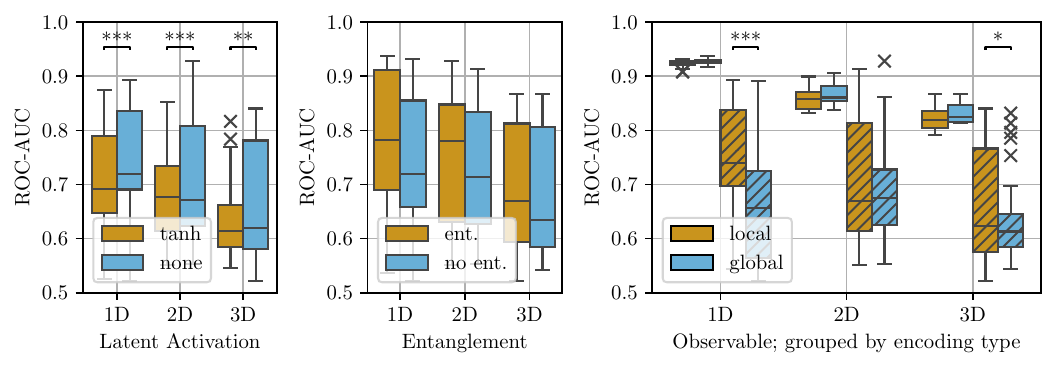}}
\caption{\textbf{Influence of quantum-specific hyperparameters on hybrid model performance}. The impact of latent layer activation function, effect of entanglement and type of observable is statistically compared for hybrid models (*: $p<0.05$, **: $p<0.01$, ***: $p<0.001$). In the observable comparison, striped boxplots represent angle encoding models, while non-striped boxplots represent amplitude encoding models. For amplitude encoding, this observable comparison includes only Amp-Gen, since QCNN uses a single local measurement by design.}
\label{fig:hybrid_hparams_boxplot}
\end{figure}

The choice of observable yielded different results depending on the encoding scheme (Fig.~\ref{fig:hybrid_hparams_boxplot}, right panel). For angle encoding models, using \textbf{local} Pauli-Z observables resulted in statistically significantly better median performance on 1D and 3D datasets compared to a \textbf{global} observable, with overall larger interquartile range. For amplitude encoding models (note: here only \textbf{Amp-Gen}, since \textbf{QCNN} uses a single local measurement by design), there was a nonsignificant trend favoring the \textbf{global} observable with higher median scores. Table~\ref{tab:results} summarizes the performance across key metrics, comparing classical models against hybrid models grouped by QNN type.

\begin{table}[t!]
\centering
\caption{\textbf{Performance metrics for classical and hybrid models.} Classical models are grouped together, hybrid models are split by QNN type. For each group the median score achieved is shown as well as minima and maxima in brackets. Bold highlights best median scores within each data modality.}
\label{tab:results}
\begin{tabular}{|c|l|c|c|c|}
\hline
\textbf{Dimension} & \textbf{Configuration} & \textbf{ROC-AUC} & \textbf{Avg. Precision} & \textbf{Balanced Acc.}\\
\hline
& Classical & $0.91_{[0.83, 0.94]}$ & $0.90_{[0.81, 0.93]}$ & $0.84_{[0.77, 0.86]}$ \\
& Ang-RY & $0.80_{[0.53, 0.89]}$ & $0.82_{[0.56, 0.90]}$ & $0.70_{[0.50, 0.83]}$ \\
1D & Ang-Arb & $0.69_{[0.52, 0.79]}$ & $0.70_{[0.55, 0.77]}$ & $0.59_{[0.50, 0.70]}$ \\
& Amp-Gen & $\mathbf{0.93_{[0.91, 0.94]}}$ & $\mathbf{0.92_{[0.90, 0.94]}}$ & $\mathbf{0.85_{[0.74, 0.87]}}$ \\
& QCNN & $0.92_{[0.92, 0.93]}$ & $0.92_{[0.91, 0.92]}$ & $0.79_{[0.78, 0.82]}$ \\
\hline\hline
& Classical & $\mathbf{0.87_{[0.83, 0.94]}}$ & $0.94_{[0.91, 0.97]}$ & $\mathbf{0.83_{[0.77, 0.86]}}$ \\
& Ang-RY & $0.79_{[0.66, 0.93]}$ & $0.92_{[0.85, 0.97]}$ & $0.71_{[0.50, 0.86]}$ \\
2D & Ang-Arb & $0.62_{[0.55, 0.69]}$ & $0.82_{[0.77, 0.87]}$ & $0.55_{[0.50, 0.65]}$ \\
& Amp-Gen & $0.86_{[0.83, 0.91]}$ & $0.94_{[0.93, 0.96]}$ & $0.78_{[0.74, 0.83]}$ \\
& QCNN & $0.87_{[0.84, 0.90]}$ & $\mathbf{0.95_{[0.93, 0.96]}}$ & $0.77_{[0.75, 0.80]}$ \\
\hline\hline
& Classical & $\mathbf{0.87_{[0.83, 0.90]}}$ & $\mathbf{0.78_{[0.75, 0.81]}}$ & $\mathbf{0.79_{[0.74, 0.81]}}$ \\
& Ang-RY & $0.70_{[0.60, 0.84]}$ & $0.51_{[0.37, 0.74]}$ & $0.55_{[0.50, 0.73]}$ \\
3D & Ang-Arb & $0.58_{[0.52, 0.64]}$ & $0.34_{[0.30, 0.40]}$ & $0.51_{[0.50, 0.53]}$ \\
& Amp-Gen & $0.82_{[0.79, 0.87]}$ & $0.71_{[0.68, 0.78]}$ & $0.70_{[0.54, 0.75]}$ \\
& QCNN & $0.83_{[0.82, 0.87]}$ & $0.72_{[0.71, 0.77]}$ & $0.63_{[0.61, 0.68]}$ \\
\hline
\end{tabular}
\end{table}
%
%
%
\section{Discussion}
\label{sec:discussion}
Hybrid quantum-classical neural networks (HQNNs) have gained considerable attention as potential near-term applications of quantum computing, particularly for processing high-dimensional data. Yet the specific contribution of their quantum components is often overlooked, with emphasis typically placed on overall model performance. From our perspective, a more granular understanding of quantum contributions is crucial for advancing hybrid QML in a meaningful way. Motivated by this gap, our study systematically assessed the role of quantum processing within selected HQNN architectures across three real-world medical datasets spanning 1D ECG signals, 2D breast ultrasound images, and 3D chest CT scans. The empirical findings offer a nuanced perspective on the contribution of the tested quantum components to overall model performance under ideal, noise-free simulation.

Our results revealed that while some hybrid model configurations achieved scores comparable to matched classical counterparts, in most cases the hybrid models analyzed in the study underperformed relative to their classical references. This observation aligns with recent studies that found limited empirical contributions of quantum components in specific hybrid quantum-classical configurations \cite{bowles_better_2024,kolle_disentangling_2024}. Most of the investigated hybrid models exhibited lower median performance and greater variability, indicating that, within the tested setting, simply incorporating quantum components into classical architectures does not inherently improve their performance and may even hinder it under certain conditions.

\subsection{Influence of Classical Pre-Processing Layers}
The extent of classical pre-processing had a significant effect on the performance of the investigated hybrid models. In the 2D and 3D datasets, performance improved with deeper classical feature extractors (\textbf{3conv} and \textbf{1conv}). While this trend was also observable in purely classical models, it was especially relevant for hybrid models, suggesting that, in the tested architectures, performance depended strongly on the amount of preceding classical feature extraction. The improved performance with deeper classical layers in higher-dimensional data (2D, 3D) suggests a greater reliance on classical pre-processing. However, as dimensionality does not equate to classification complexity, we avoid cross-dataset comparisons.

\subsection{Influence of Latent Space Dimension}
We further observed that hybrid models benefited from higher classical compression, i.e., smaller latent spaces. This was particularly evident in the 1D dataset, where \textbf{256} latent features encoded into the quantum circuit led to significantly worse results than encoding just \textbf{16} features. Classical models, in contrast, were less sensitive to latent space size, suggesting that the tested QNN configurations may be less effective at exploiting larger latent representations \cite{periyasamy_incremental_2022}. These findings suggest either that increased classical processing improves performance (implying minimal quantum contribution), or that shallower quantum circuits learn more effectively, making overparameterization in quantum models less advantageous. Although we did not directly measure gradient variances in our experiments, one possible theoretical explanation is suggested by barren-plateau results, which predict that gradient variance decays exponentially with both circuit depth and qubit count, leading to increased training difficulty for deeper models \cite{mcclean_barren_2018}.

\subsection{QNN Architecture and Feature Encoding}
Among the quantum-specific hyperparameters, the choice of encoding method had the most substantial impact. Across all modalities, amplitude encoding significantly outperformed angle encoding, a result consistent with prior studies on hybrid models \cite{chen_hybrid_2025}. A possible explanation is given by \cite{schuld_effect_2021}, who prove that quantum models with angle encoding correspond to low‐order Fourier series whose frequencies are determined by each single data upload, resulting in a restricted function class when features are uploaded only once in the circuit. To isolate the effect of the encoding scheme, we kept the number of qubits constant across both amplitude and angle encoding architectures. However, since angle encoding requires a number of encoding gates that scales linearly with the number of features (unlike the logarithmic scaling of amplitude encoding), we introduced additional encoding gates along the circuit to accommodate the full feature vector, as described in Section~\ref{sec:model-architecture}. This led to slightly increased circuit depth and complexity for angle encoding models, even though the number of qubits remained the same, which poses another possible reason for the worse performance of angle encoding models. Also, the incremental encoding scheme employed in both angle encoding models (\textbf{Ang-RY} and \textbf{Ang-Arb}) has led to mediocre results in earlier studies \cite{periyasamy_incremental_2022}. It should be noted, however, that amplitude encoding often requires complex state-preparation routines whose depth grows (in the worst case) exponentially with input dimension. As a result, the observed advantage may be offset by prohibitive circuit depth and gate counts on real NISQ hardware.

\subsection{Latent Layer Activation Function}
In angle encoding models, applying a $\tanh$ activation function (scaled to $[-\pi, \pi]$) prior to quantum encoding often led to worse performance compared to omitting activation altogether. While scaled activations are typically used to prevent ambiguity due to the periodic nature of rotation gates, they may also compress feature magnitudes too drastically, resulting in information loss. In contrast, allowing unbounded linear activations may risk exceeding the gate’s expected input range, but in practice, preceding classical layers appear to adapt to these constraints, a behavior observed previously \cite{lloyd_quantum_2020}.

\subsection{Entanglement in the Variational Circuit}
The inclusion of entanglement in the quantum circuits led to slightly higher median scores, although these differences were not statistically significant. This is reflected in contrasting literature: While some studies found that entanglement enhances expressivity and accuracy~\cite{nakhl_calibrating_2024}, others reported that entangled variants do not necessarily outperform their disentangled counterparts~\cite{bowles_better_2024}, leaving the precise role of entanglement in HQNNs as an important open question for future research. Importantly, our findings are limited to the fixed, generic entanglement patterns investigated here and do not rule out potential benefits from structured, problem-aware entanglement strategies.

\subsection{Type of Observable}
Local observables outperformed global observables predominantly within angle-encoding models. Conversely, amplitude-encoding models benefited more from global observables. A possible explanation is that amplitude encoding generates complex multi-qubit interactions to prepare a desired state which might be better captured by global observables. Additionally, prior work has found global observables to be more susceptible to barren plateaus \cite{cerezo_cost_2021}, owing to vanishing gradients. However, our study did not directly measure gradient variances, so this remains a theoretical consideration.

\subsection{Limitations}
While our study investigated a selection of representative HQNN architectures and quantum-specific hyperparameters, the design space of hybrid models remains vast. A comprehensive understanding of quantum contributions will require exploring an even wider spectrum of circuit topologies, integration strategies, entanglement structures and optimization methods beyond gradient-based training. In particular, we did not evaluate alternative hybrid paradigms such as quantum transfer learning with frozen classical backbones, explicit data re-uploading circuits, or problem-aware and data-dependent entanglement strategies. Moreover, experiments used ideal, noise-free simulations, a simplification favoring quantum components, whereas real hardware noise and state-preparation overheads may introduce additional challenges, especially for amplitude encoding. We also limited our analysis to binary classification tasks from medical datasets, whereas quantum and hybrid models may behave differently in alternative settings such as regression problems, unsupervised and reinforcement learning, or generative tasks. Finally, while barren-plateau theory provides a plausible explanation for some observed trends, we did not directly measure gradient variances.

\section{Conclusion}
\label{sec:conclusion}
In conclusion, this empirical assessment contributes to the quantum machine learning discourse by providing insights into the role of quantum components within the investigated hybrid models. While hybrid quantum-classical neural networks provide an attractive route to processing complex, real-world data, our results advocate for realistic expectations regarding quantum utility and careful architectural decisions, as the observed benefit of quantum components remains nuanced and highly design- and data-dependent. Nonetheless, our study marks a meaningful empirical step toward a more grounded assessment of potential quantum utility, helping to bridge the gap between theoretical promise and applied evaluation. As quantum hardware and algorithms continue to evolve, future research building on these insights will be essential for objectively evaluating quantum computing’s role in machine learning and artificial intelligence.

\begin{credits}
\subsubsection{\ackname} This work is financed by research subsidies granted by the government of Upper Austria through the projects MIMAS.ai and MIMAS.ng, and by the Austrian Research Promotion Agency (FFG) through the project QML4Med (grant no. 913256). www.ffg.at. RISC Software GmbH is a member of UAR (Upper Austrian Research) Innovation Network.

\smallskip
The authors acknowledge the use of ChatGPT, a large language model developed by OpenAI, for assistance in proofreading and improving the manuscript's grammar and clarity.

\subsubsection{\discintname}
The authors have no competing interests to declare that are
relevant to the content of this article.
\end{credits}
%
%
%
\bibliographystyle{splncs04}
\bibliography{references}

@article{cerezo_challenges_2022,
	title = {Challenges and opportunities in quantum machine learning},
	volume = {2},
	copyright = {2022 Springer Nature America, Inc.},
	issn = {2662-8457},
	url = {https://www.nature.com/articles/s43588-022-00311-3},
	doi = {10.1038/s43588-022-00311-3},
	language = {en},
	number = {9},
	urldate = {2025-04-09},
	journal = {Nature Computational Science},
	author = {Cerezo, M. and Verdon, Guillaume and Huang, Hsin-Yuan and Cincio, Lukasz and Coles, Patrick J.},
	month = sep,
	year = {2022},
	note = {Publisher: Nature Publishing Group},
	pages = {567--576},
}

@article{wang_comprehensive_2024,
	title = {A comprehensive review of quantum machine learning: from {NISQ} to fault tolerance},
	volume = {87},
	issn = {0034-4885},
	shorttitle = {A comprehensive review of quantum machine learning},
	url = {https://dx.doi.org/10.1088/1361-6633/ad7f69},
	doi = {10.1088/1361-6633/ad7f69},
	language = {en},
	number = {11},
	urldate = {2025-04-09},
	journal = {Reports on Progress in Physics},
	author = {Wang, Yunfei and Liu, Junyu},
	month = oct,
	year = {2024},
	note = {Publisher: IOP Publishing},
	pages = {116402},
}

@article{mitarai_quantum_2018,
	title = {Quantum circuit learning},
	volume = {98},
	url = {https://link.aps.org/doi/10.1103/PhysRevA.98.032309},
	doi = {10.1103/PhysRevA.98.032309},
	number = {3},
	urldate = {2025-04-11},
	journal = {Physical Review A},
	author = {Mitarai, K. and Negoro, M. and Kitagawa, M. and Fujii, K.},
	month = sep,
	year = {2018},
	note = {Publisher: American Physical Society},
	pages = {032309},
}

@article{sharma_noise_2020,
	title = {Noise resilience of variational quantum compiling},
	volume = {22},
	issn = {1367-2630},
	url = {https://dx.doi.org/10.1088/1367-2630/ab784c},
	doi = {10.1088/1367-2630/ab784c},
	language = {en},
	number = {4},
	urldate = {2025-04-09},
	journal = {New Journal of Physics},
	author = {Sharma, Kunal and Khatri, Sumeet and Cerezo, M and Coles, Patrick J},
	month = apr,
	year = {2020},
	note = {Publisher: IOP Publishing},
	pages = {043006},
}

@article{abbas_power_2021,
	title = {The power of quantum neural networks},
	volume = {1},
	copyright = {2021 The Author(s), under exclusive licence to Springer Nature America, Inc.},
	issn = {2662-8457},
	url = {https://www.nature.com/articles/s43588-021-00084-1},
	doi = {10.1038/s43588-021-00084-1},
	language = {en},
	number = {6},
	urldate = {2025-04-11},
	journal = {Nature Computational Science},
	author = {Abbas, Amira and Sutter, David and Zoufal, Christa and Lucchi, Aurelien and Figalli, Alessio and Woerner, Stefan},
	month = jun,
	year = {2021},
	note = {Publisher: Nature Publishing Group},
	pages = {403--409},
}

@misc{bowles_better_2024,
	title = {Better than classical? {The} subtle art of benchmarking quantum machine learning models},
	url = {http://arxiv.org/abs/2403.07059},
	doi = {10.48550/arXiv.2403.07059},
	urldate = {2025-03-12},
	publisher = {arXiv},
	author = {Bowles, Joseph and Ahmed, Shahnawaz and Schuld, Maria},
	month = mar,
	year = {2024},
	note = {arXiv:2403.07059 [quant-ph]},
}

@misc{kolle_disentangling_2024,
	title = {Disentangling {Quantum} and {Classical} {Contributions} in {Hybrid} {Quantum} {Machine} {Learning} {Architectures}},
	url = {http://arxiv.org/abs/2311.05559},
	doi = {10.48550/arXiv.2311.05559},
	urldate = {2025-04-11},
	publisher = {arXiv},
	author = {Kölle, Michael and Maurer, Jonas and Altmann, Philipp and Sünkel, Leo and Stein, Jonas and Linnhoff-Popien, Claudia},
	month = jan,
	year = {2024},
	note = {arXiv:2311.05559 [quant-ph]},
}

@article{mari_transfer_2020,
	title = {Transfer learning in hybrid classical-quantum neural networks},
	volume = {4},
	issn = {2521-327X},
	url = {http://arxiv.org/abs/1912.08278},
	doi = {10.22331/q-2020-10-09-340},
	urldate = {2025-03-17},
	journal = {Quantum},
	author = {Mari, Andrea and Bromley, Thomas R. and Izaac, Josh and Schuld, Maria and Killoran, Nathan},
	month = oct,
	year = {2020},
	note = {arXiv:1912.08278 [quant-ph]},
	pages = {340},
}

@article{goldberger_physiobank_2000,
	title = {{PhysioBank}, {PhysioToolkit}, and {PhysioNet}: components of a new research resource for complex physiologic signals},
	volume = {101},
	issn = {1524-4539},
	shorttitle = {{PhysioBank}, {PhysioToolkit}, and {PhysioNet}},
	doi = {10.1161/01.cir.101.23.e215},
	language = {eng},
	number = {23},
	journal = {Circulation},
	author = {Goldberger, A. L. and Amaral, L. A. and Glass, L. and Hausdorff, J. M. and Ivanov, P. C. and Mark, R. G. and Mietus, J. E. and Moody, G. B. and Peng, C. K. and Stanley, H. E.},
	month = jun,
	year = {2000},
	pmid = {10851218},
	pages = {E215--220},
}

@article{moody_impact_2001,
	title = {The impact of the {MIT}-{BIH} {Arrhythmia} {Database}},
	volume = {20},
	issn = {1937-4186},
	doi = {10.1109/51.932724},
	number = {3},
	journal = {IEEE Engineering in Medicine and Biology Magazine},
	author = {Moody, G.B. and Mark, R.G.},
	month = may,
	year = {2001},
	pages = {45--50},
}

@article{yang_medmnist_2023,
	title = {{MedMNIST} v2 - {A} large-scale lightweight benchmark for {2D} and {3D} biomedical image classification},
	volume = {10},
	copyright = {2023 The Author(s)},
	issn = {2052-4463},
	url = {https://www.nature.com/articles/s41597-022-01721-8},
	doi = {10.1038/s41597-022-01721-8},
	language = {en},
	number = {1},
	urldate = {2025-03-17},
	journal = {Scientific Data},
	author = {Yang, Jiancheng and Shi, Rui and Wei, Donglai and Liu, Zequan and Zhao, Lin and Ke, Bilian and Pfister, Hanspeter and Ni, Bingbing},
	month = jan,
	year = {2023},
	note = {Publisher: Nature Publishing Group},
	pages = {41},
}

@article{schuld_effect_2021,
	title = {The effect of data encoding on the expressive power of variational quantum machine learning models},
	volume = {103},
	issn = {2469-9926, 2469-9934},
	url = {http://arxiv.org/abs/2008.08605},
	doi = {10.1103/PhysRevA.103.032430},
	number = {3},
	urldate = {2025-02-26},
	journal = {Physical Review A},
	author = {Schuld, Maria and Sweke, Ryan and Meyer, Johannes Jakob},
	month = mar,
	year = {2021},
	note = {arXiv:2008.08605 [quant-ph]},
	pages = {032430},
}

@misc{bergholm_pennylane_2022,
	title = {{PennyLane}: {Automatic} differentiation of hybrid quantum-classical computations},
	shorttitle = {{PennyLane}},
	url = {http://arxiv.org/abs/1811.04968},
	doi = {10.48550/arXiv.1811.04968},
	urldate = {2025-03-21},
	publisher = {arXiv},
	author = {Bergholm, Ville and Izaac, Josh and Schuld, Maria and Gogolin, Christian and Ahmed, Shahnawaz and Ajith, Vishnu and Alam, M. Sohaib and Alonso-Linaje, Guillermo and AkashNarayanan, B. and Asadi, Ali and Arrazola, Juan Miguel and Azad, Utkarsh and Banning, Sam and Blank, Carsten and Bromley, Thomas R. and Cordier, Benjamin A. and Ceroni, Jack and Delgado, Alain and Matteo, Olivia Di and Dusko, Amintor and Garg, Tanya and Guala, Diego and Hayes, Anthony and Hill, Ryan and Ijaz, Aroosa and Isacsson, Theodor and Ittah, David and Jahangiri, Soran and Jain, Prateek and Jiang, Edward and Khandelwal, Ankit and Kottmann, Korbinian and Lang, Robert A. and Lee, Christina and Loke, Thomas and Lowe, Angus and McKiernan, Keri and Meyer, Johannes Jakob and Montañez-Barrera, J. A. and Moyard, Romain and Niu, Zeyue and O'Riordan, Lee James and Oud, Steven and Panigrahi, Ashish and Park, Chae-Yeun and Polatajko, Daniel and Quesada, Nicolás and Roberts, Chase and Sá, Nahum and Schoch, Isidor and Shi, Borun and Shu, Shuli and Sim, Sukin and Singh, Arshpreet and Strandberg, Ingrid and Soni, Jay and Száva, Antal and Thabet, Slimane and Vargas-Hernández, Rodrigo A. and Vincent, Trevor and Vitucci, Nicola and Weber, Maurice and Wierichs, David and Wiersema, Roeland and Willmann, Moritz and Wong, Vincent and Zhang, Shaoming and Killoran, Nathan},
	month = jul,
	year = {2022},
	note = {arXiv:1811.04968 [quant-ph]},
}

@misc{paszke_pytorch_2019,
	title = {{PyTorch}: {An} {Imperative} {Style}, {High}-{Performance} {Deep} {Learning} {Library}},
	shorttitle = {{PyTorch}},
	url = {http://arxiv.org/abs/1912.01703},
	doi = {10.48550/arXiv.1912.01703},
	urldate = {2025-04-14},
	publisher = {arXiv},
	author = {Paszke, Adam and Gross, Sam and Massa, Francisco and Lerer, Adam and Bradbury, James and Chanan, Gregory and Killeen, Trevor and Lin, Zeming and Gimelshein, Natalia and Antiga, Luca and Desmaison, Alban and Köpf, Andreas and Yang, Edward and DeVito, Zach and Raison, Martin and Tejani, Alykhan and Chilamkurthy, Sasank and Steiner, Benoit and Fang, Lu and Bai, Junjie and Chintala, Soumith},
	month = dec,
	year = {2019},
	note = {arXiv:1912.01703 [cs]},
}

@article{perez-salinas_data_2020,
	title = {Data re-uploading for a universal quantum classifier},
	volume = {4},
	url = {https://quantum-journal.org/papers/q-2020-02-06-226/},
	doi = {10.22331/q-2020-02-06-226},
	language = {en-GB},
	urldate = {2025-02-25},
	journal = {Quantum},
	author = {Pérez-Salinas, Adrián and Cervera-Lierta, Alba and Gil-Fuster, Elies and Latorre, José I.},
	month = feb,
	year = {2020},
	note = {Publisher: Verein zur Förderung des Open Access Publizierens in den Quantenwissenschaften},
	pages = {226},
}

@article{zeng_multi-classification_2022,
	title = {A {Multi}-{Classification} {Hybrid} {Quantum} {Neural} {Network} {Using} an {All}-{Qubit} {Multi}-{Observable} {Measurement} {Strategy}},
	volume = {24},
	copyright = {http://creativecommons.org/licenses/by/3.0/},
	issn = {1099-4300},
	url = {https://www.mdpi.com/1099-4300/24/3/394},
	doi = {10.3390/e24030394},
	language = {en},
	number = {3},
	urldate = {2025-03-21},
	journal = {Entropy},
	author = {Zeng, Yi and Wang, Hao and He, Jin and Huang, Qijun and Chang, Sheng},
	month = mar,
	year = {2022},
	note = {Number: 3
Publisher: Multidisciplinary Digital Publishing Institute},
	pages = {394},
}

@article{hafeez_h-qnn_2024,
	title = {H-{QNN}: {A} {Hybrid} {Quantum}–{Classical} {Neural} {Network} for {Improved} {Binary} {Image} {Classification}},
	volume = {5},
	copyright = {http://creativecommons.org/licenses/by/3.0/},
	issn = {2673-2688},
	shorttitle = {H-{QNN}},
	url = {https://www.mdpi.com/2673-2688/5/3/70},
	doi = {10.3390/ai5030070},
	language = {en},
	number = {3},
	urldate = {2025-03-21},
	journal = {AI},
	author = {Hafeez, Muhammad Asfand and Munir, Arslan and Ullah, Hayat},
	month = sep,
	year = {2024},
	note = {Number: 3
Publisher: Multidisciplinary Digital Publishing Institute},
	pages = {1462--1481},
}

@article{sagingalieva_hybrid_2023,
	title = {Hybrid {Quantum} {Neural} {Network} for {Drug} {Response} {Prediction}},
	volume = {15},
	copyright = {http://creativecommons.org/licenses/by/3.0/},
	issn = {2072-6694},
	url = {https://www.mdpi.com/2072-6694/15/10/2705},
	doi = {10.3390/cancers15102705},
	language = {en},
	number = {10},
	urldate = {2025-03-10},
	journal = {Cancers},
	author = {Sagingalieva, Asel and Kordzanganeh, Mohammad and Kenbayev, Nurbolat and Kosichkina, Daria and Tomashuk, Tatiana and Melnikov, Alexey},
	month = jan,
	year = {2023},
	note = {Number: 10
Publisher: Multidisciplinary Digital Publishing Institute},
	pages = {2705},
}

@misc{perelshtein_practical_2022,
	title = {Practical application-specific advantage through hybrid quantum computing},
	url = {http://arxiv.org/abs/2205.04858},
	doi = {10.48550/arXiv.2205.04858},
	urldate = {2025-03-10},
	publisher = {arXiv},
	author = {Perelshtein, Michael and Sagingalieva, Asel and Pinto, Karan and Shete, Vishal and Pakhomchik, Alexey and Melnikov, Artem and Neukart, Florian and Gesek, Georg and Melnikov, Alexey and Vinokur, Valerii},
	month = may,
	year = {2022},
	note = {arXiv:2205.04858 [quant-ph]},
}

@article{rainjonneau_quantum_2023,
	title = {Quantum {Algorithms} {Applied} to {Satellite} {Mission} {Planning} for {Earth} {Observation}},
	volume = {16},
	issn = {2151-1535},
	url = {https://ieeexplore.ieee.org/document/10155128},
	doi = {10.1109/JSTARS.2023.3287154},
	urldate = {2025-03-10},
	journal = {IEEE Journal of Selected Topics in Applied Earth Observations and Remote Sensing},
	author = {Rainjonneau, Serge and Tokarev, Igor and Iudin, Sergei and Rayaprolu, Saaketh and Pinto, Karan and Lemtiuzhnikova, Daria and Koblan, Miras and Barashov, Egor and Kordzanganeh, Mo and Pflitsch, Markus and Melnikov, Alexey},
	year = {2023},
	note = {Conference Name: IEEE Journal of Selected Topics in Applied Earth Observations and Remote Sensing},
	pages = {7062--7075},
}

@article{xiang_quantum_2024,
	title = {Quantum classical hybrid convolutional neural networks for breast cancer diagnosis},
	volume = {14},
	copyright = {2024 The Author(s)},
	issn = {2045-2322},
	url = {https://www.nature.com/articles/s41598-024-74778-7},
	doi = {10.1038/s41598-024-74778-7},
	language = {en},
	number = {1},
	urldate = {2025-03-21},
	journal = {Scientific Reports},
	author = {Xiang, Qiuyu and Li, Dongfen and Hu, Zhikang and Yuan, Yuhang and Sun, Yuchen and Zhu, Yonghao and Fu, You and Jiang, Yangyang and Hua, Xiaoyu},
	month = oct,
	year = {2024},
	note = {Publisher: Nature Publishing Group},
	pages = {24699},
}

@inproceedings{matic_quantum-classical_2022,
	title = {Quantum-classical convolutional neural networks in radiological image classification},
	url = {https://ieeexplore.ieee.org/document/9951255},
	doi = {10.1109/QCE53715.2022.00024},
	urldate = {2025-03-21},
	booktitle = {2022 {IEEE} {International} {Conference} on {Quantum} {Computing} and {Engineering} ({QCE})},
	author = {Matic, Andrea and Monnet, Maureen and Lorenz, Jeanette Miriam and Schachtner, Balthasar and Messerer, Thomas},
	month = sep,
	year = {2022},
	pages = {56--66},
}

@article{cong_quantum_2019,
	title = {Quantum convolutional neural networks},
	volume = {15},
	copyright = {2019 The Author(s), under exclusive licence to Springer Nature Limited},
	issn = {1745-2481},
	url = {https://www.nature.com/articles/s41567-019-0648-8},
	doi = {10.1038/s41567-019-0648-8},
	language = {en},
	number = {12},
	urldate = {2025-04-11},
	journal = {Nature Physics},
	author = {Cong, Iris and Choi, Soonwon and Lukin, Mikhail D.},
	month = dec,
	year = {2019},
	note = {Publisher: Nature Publishing Group},
	pages = {1273--1278},
}

@misc{qiskit_convnet,
  author       = {{Qiskit Community}},
  title        = {{Quantum Convolutional Neural Networks Tutorial}},
  howpublished = {\url{https://qiskit-community.github.io/qiskit-machine-learning/tutorials/11_quantum_convolutional_neural_networks.html}},
  note         = {[Online; Accessed: Apr. 11, 2025]}
}

@article{mcclean_barren_2018,
	title = {Barren plateaus in quantum neural network training landscapes},
	volume = {9},
	copyright = {2018 The Author(s)},
	issn = {2041-1723},
	url = {https://www.nature.com/articles/s41467-018-07090-4},
	doi = {10.1038/s41467-018-07090-4},
	language = {en},
	number = {1},
	urldate = {2025-04-10},
	journal = {Nature Communications},
	author = {McClean, Jarrod R. and Boixo, Sergio and Smelyanskiy, Vadim N. and Babbush, Ryan and Neven, Hartmut},
	month = nov,
	year = {2018},
	note = {Publisher: Nature Publishing Group},
	pages = {4812},
}

@misc{chen_hybrid_2025,
	title = {Hybrid {Quantum} {Neural} {Networks} with {Amplitude} {Encoding}: {Advancing} {Recovery} {Rate} {Predictions}},
	shorttitle = {Hybrid {Quantum} {Neural} {Networks} with {Amplitude} {Encoding}},
	url = {http://arxiv.org/abs/2501.15828},
	doi = {10.48550/arXiv.2501.15828},
	urldate = {2025-04-10},
	publisher = {arXiv},
	author = {Chen, Ying and Griffin, Paul and Recchia, Paolo and Zhou, Lei and Zhang, Hongrui},
	month = feb,
	year = {2025},
	note = {arXiv:2501.15828 [q-fin]},
}

@article{nakhl_calibrating_2024,
	title = {Calibrating the role of entanglement in variational quantum circuits},
	volume = {109},
	issn = {2469-9926, 2469-9934},
	url = {http://arxiv.org/abs/2310.10885},
	doi = {10.1103/PhysRevA.109.032413},
	number = {3},
	urldate = {2025-04-10},
	journal = {Physical Review A},
	author = {Nakhl, Azar C. and Quella, Thomas and Usman, Muhammad},
	month = mar,
	year = {2024},
	note = {arXiv:2310.10885 [quant-ph]},
	pages = {032413},
}

@misc{lloyd_quantum_2020,
	title = {Quantum embeddings for machine learning},
	url = {http://arxiv.org/abs/2001.03622},
	doi = {10.48550/arXiv.2001.03622},
	urldate = {2025-04-10},
	publisher = {arXiv},
	author = {Lloyd, Seth and Schuld, Maria and Ijaz, Aroosa and Izaac, Josh and Killoran, Nathan},
	month = feb,
	year = {2020},
	note = {arXiv:2001.03622 [quant-ph]},
}

@article{cerezo_cost_2021,
	title = {Cost function dependent barren plateaus in shallow parametrized quantum circuits},
	volume = {12},
	copyright = {2021 The Author(s)},
	issn = {2041-1723},
	url = {https://www.nature.com/articles/s41467-021-21728-w},
	doi = {10.1038/s41467-021-21728-w},
	language = {en},
	number = {1},
	urldate = {2025-04-11},
	journal = {Nature Communications},
	author = {Cerezo, M. and Sone, Akira and Volkoff, Tyler and Cincio, Lukasz and Coles, Patrick J.},
	month = mar,
	year = {2021},
	note = {Publisher: Nature Publishing Group},
	pages = {1791},
}

@inproceedings{periyasamy_incremental_2022,
	title = {Incremental {Data}-{Uploading} for {Full}-{Quantum} {Classification}},
	url = {http://arxiv.org/abs/2205.03057},
	doi = {10.1109/QCE53715.2022.00021},
	urldate = {2025-04-11},
	booktitle = {2022 {IEEE} {International} {Conference} on {Quantum} {Computing} and {Engineering} ({QCE})},
	author = {Periyasamy, Maniraman and Meyer, Nico and Ufrecht, Christian and Scherer, Daniel D. and Plinge, Axel and Mutschler, Christopher},
	month = sep,
	year = {2022},
	note = {arXiv:2205.03057 [quant-ph]},
	pages = {31--37},
}

\end{document}